\documentclass[reprint, amsmath,amssymb, amsfonts, aps,prd]{revtex4-1}
\pdfoutput=1
\usepackage{natbib,hyperref,ifthen} 
\usepackage[vcentermath,noautoscale,stdtext]{youngtab}
\usepackage[vcentermath,noautoscale,stdtext]{youngtab}
\usepackage{varioref,exscale,latexsym,amsmath,amssymb}
\usepackage[pdftex]{graphicx}
\usepackage[left=3cm, right=3cm, top=2cm, bottom=2.75cm]{geometry} 
\usepackage{dcolumn}
\usepackage{bm}

\newcommand{\Tr}{\mathrm{Tr}}
\newcommand{\be}{\begin{equation}}
\newcommand{\ee}{\end{equation}}

\newcommand{\n}{\nonumber \\}
\newcommand{\eqn}[1]{Eq.~(\ref{#1})}

\begin{document}

\title{Lattice-Friendly Gauge Completion of a Composite Higgs  with \\ Top Partners}
\author{Helene Gertov$^{1,2}$}\email{gertov@cp3-origins.net}
\author{Ann E.~Nelson$^2$}\email{aenelson@uw.edu}
\author{Ashley Perko$^3$}\email{perko@dartmouth.edu}
\author{Devin G.~E.~Walker$^3$}\email{devin.g.walker@dartmouth.edu}
\affiliation{$^1$CP$^3$-Origins, University of Southern Denmark, Campusvej 55, 5230 Odense, Denmark}
\affiliation{$^2$Department of Physics, Box 1560, University of Washington, Seattle, WA 98195-1560 USA}
\affiliation{$^3$Department of Physics and Astronomy, Dartmouth College, Hanover, NH 03755 USA}

\begin{abstract}
We give an explicit example of a composite Higgs model  with a pseudo-Nambu-Goldstone Higgs in which the top Yukawa coupling is generated via the partial compositeness mechanism. This mechanism requires composite top partners which are relatively light compared to the typical mass scale of the strongly coupled theory. While most studies of the phenomenology of such models have focused on a bottom-up approach with a minimal effective theory, a top-down approach suggests that the theory should contain a limit in which an unbroken  global chiral symmetry protects the mass of the   top partners, and the spectrum of the partners satisfies `t~Hooft matching conditions.   We find  that the  relatively light fermions and pseudo-Goldstone bosons  fall into complete multiplets of a large approximate global symmetry, and that the spectrum of  particles lighter than a few TeV is non-minimal.  Our example illustrates the  likely features of a such a composite Higgs theory   and also serves as an example of a non-chiral theory   with a possible solution to the `t~Hooft matching conditions.    We find in this example that for some low-energy parameters in the effective theory the top partners can decay into high-multiplicity final states, which could be difficult for the Large Hadron Collider (LHC) to constrain. This may potentially allow for the top partners to be  lighter than those in more minimal models.
\end{abstract}

\maketitle

\section{\label{sec:intro}Introduction}

One of the most important questions facing particle physicists is understanding the mechanism of electroweak symmetry breaking. The minimal Standard Model does a great job of describing this in a manner which is in good agreement with all collider data, but theoretically the  size of the Higgs mass is puzzling.  Precision electroweak corrections  and the absence of any new particle discovery at the LHC suggests   that the Standard Model  may be a good description of particle physics up to  a cutoff of at least a few TeV.  Dimensional analysis would then give a Higgs mass   of order of this cutoff unless there is either some fine-tuning or some symmetry reason for it to be lighter.  One possible reason for a light Higgs is   that it is an approximate Nambu-Goldstone boson, whose mass is protected by an approximate non-linearly realized symmetry. However, such a symmetry must be  broken to allow non-derivative couplings.  The largest such coupling is to the top quark, and the the top Yukawa coupling leads to large corrections to the Higgs mass squared. In theories with light top partners,  such corrections can be partially canceled and the Higgs can be naturally light compared to the cutoff. 

A new   gauge theory which confines the Higgs and the top partners has to have certain dynamical features which are different from those of QCD. One non-QCD-like feature is that the theory should have  composite fermions which are light compared with the compositeness scale to serve as top partners. One possibility is a theory which has  a limit in which it confines but does not break all of the chiral symmetry, and which contains massless fermions which match the `t~Hooft anomaly conditions that may serve as top partners. A second feature is that   the theory should be near a limit in which the chiral global symmetry  which is protecting the mass of the top partners does   spontaneously break, in order to produce a pseudo-Nambu-Goldstone boson multiplet containing the composite Higgs which coupled to the top partners. In order for  a small perturbation to result  in chiral symmetry breaking, the unperturbed theory must be near a second-order phase transition, and the spectrum of the unperturbed theory would then be expected to contain a relatively light scalar which is not a Nambu-Goldstone boson.  

The first example of an ultraviolet (UV) completion of a theory with light composite top  satisfying 't Hooft anomaly matching conditions for an approximate symmetry was studied in \cite{Katz:2003sn}, which supplied a composite UV completion to the $SU(5)/SO(5)$ ``Littlest Higgs" version of a composite Higgs  model \cite{Georgi:1984af,ArkaniHamed:2002qy}. Other potential examples have been given in \cite{Cacciapaglia:2015vrx}.  The   assumption that such dynamics are possible is motivated by a recent lattice study. In \cite{Brower:2015owo}, evidence of an anomalously light scalar was found for a system of $12$ fermion flavors coupled to $SU(3)$ gauge fields. This scalar could be anomalously light by being an order parameter for a second-order phase transition. It would be interesting to see whether lattice studies could provide evidence for anomalously light fermions in some model  as well. Lattice studies of dynamics are limited to vector-like gauge theories with real  positive fermion determinant, and therefore theories that have no scalars with Yukawa couplings. In our model the confining group is a vector-like gauge theory.

\section{\label{sec:description}Description of Model}

Our composite pseudo-Nambu-Goldstone Higgs model is based on the $SU(4)/Sp(4)$ symmetry-breaking pattern \cite{Kaplan:1983sm} of the ``Intermediate" composite Higgs model \cite{Katz:2005au}.  We assume top partners which are in a multiplet of an approximate chiral global symmetry. We also assume that the chiral symmetry-breaking scale is low compared to the confinement scale,   and the global chiral anomalies are matched by composite fermions.  Provided our assumptions about the dynamics are correct,  our model can provide a realistic composite Higgs with a top Yukawa coupling and top partners. 

We need a model with a large enough global symmetry to embed the Standard Model gauge group, in order to allow for colored and electroweak-charged composite fermions,  as well as the non-linearly realized symmetry protecting the Higgs mass. The minimal model we found which also allows for a solution to the 't Hooft matching conditions has $10$ fermion flavors coupled to a symplectic gauge group, as well as an adjoint fermion. The fermion content is that of a supersymmetric theory. From previous work on supersymmetric theories, we know that the $SU(10)\times U(1)$ global symmetry is   confining with anomaly matching between the UV and the infrared (IR) confined phase when the gauge group is $Sp(6)$  \cite{Csaki:1996zb}.  Giving mass to the supersymmetric scalars does not change the anomaly matching, although it might change the dynamics in the limit where the scalars are heavy compared with the confinement scale. The $Sp(6)$   gauge group   is automatically  free  of gauge anomalies as long as there are an even number of pseudo-real representations. The fundamental fermions, which we will call preons, will be confined   at a scale $\Lambda$ into composite scalars (mesons) and fermions (baryons), which furnish the composite Higgs as well as a top partner that will mix with the fundamental top through the mechanism of partial compositeness. One possibility is to assume the supersymmetry of this theory is broken and the scalar superpartners have mass   below the compositeness scale. This would allow us to perform a systematic analytic understanding of the dynamics, but  simulating the the theory on the lattice would suffer from a sign problem and would not be feasible. In this work we assume the supersymmetric scalars are at or well above the compositeness scale or are even absent. 

According to  the Vafa-Witten theorem \cite{Vafa:1983tf}, in vector-like gauge theories with no scalars coupled to the fermions,    there can be no massless composites with massive constituents, ie. there can be no  violation of the persistent mass condition \cite{Preskill:1981sr}. Our composite top partners violate this condition as their mass is protected by the SU(10) global chiral symmetry, but they contain a fermion adjoint which can be given a mass without violating the symmetry. We therefore expect that in the highly non-supersymmetric limit in which the scalar masses which are   heavy compared to the compositeness scale, the chiral global symmetries of the theory will spontaneously break to a subgroup which allows fermion masses. We speculate that the chiral symmetry-breaking scale and the fermion masses in this case could be below the compositeness scale, which would imply that the chiral symmetry-breaking must be describable by a linear sigma model containing composite scalars which are anomalously light compared to the compositeness scale as well. 

In order to produce a pseudo-Goldstone Higgs and masses for the top partners,  the $SU(10)$ symmetry must spontaneously break. We assume this breaking scale is lower than the compositeness scale, which requires that even in the absence of symmetry breaking, the theory contains a relatively light scalar multiplet with a non-trivial $SU(10)$ transformation. We assume that a relatively small mass for the scalar arises because the theory with this matter content is near a second-order phase transition.  Because this is a non-supersymmetric, vector-like theory, this  assumption  can be checked on the lattice. The $SU(10)\times U(1)$ global symmetry will also be explicitly broken by preon mass terms and Standard Model gauge interactions.  

It is convenient to analyze the particle content under an $SU(6) \times SU(4)\times U(1)$ subgroup.  The $SU(4)$ will contain the electroweak and custodial symmetries, and the $SU(6)$ will contain the color group in the diagonal subgroup of $SU(3) \times SU(3) \subset SU(6)$. When an effective theory for the light mesons and baryons of this theory is analyzed, we will find that for some parameters we can trigger spontaneous breaking of the $SU(4)$ to $Sp(4)$ at a lower scale $f$. This coset structure was explored in \cite{Katz:2005au}, and is known to be the minimal coset that can form a pseudo-Goldstone composite Higgs which admits a fermionic UV completion. UV completions of this coset have previously been studied in \cite{ Barnard:2013zea, Ferretti:2013kya, Ferretti:2014qta, Cacciapaglia:2014uja, Cacciapaglia:2015eqa, Vecchi:2015fma, Sannino:2016sfx}. A $SU(6) \times SU(4)$ global symmetry structure was previously studied in the context of a $Sp(6)$ gauge group in \cite{Cacciapaglia:2015eqa}, but this work did not give a dynamical explanation for the mass and couplings of the top partner, which is provided here by the $SU(10)$ breaking. Furthermore, that work assumed a different condensate structure that breaks the global $SU(6)$ to $SO(6)$, rather than the $Sp(6)$ that we will find, as well as having the composite fermions in a different representation of the global $SU(6) \times SU(4)$.

\section{\label{sec:confine} UV Theory}

In the UV, we consider a model with a new $Sp(6)$ gauge group that confines at the scale $\Lambda$, which will be of order 10 TeV.  The matter content above $\Lambda$ consists of fermions $F$ in the fundamental representation of $Sp(6)$, which have a global flavor symmetry in the fundamental of $SU(10)$, as well as $A$, a fermionic adjoint of $Sp(6)$ that is a singlet under the flavor symmetry and is needed to form the composite fermions. There is also an anomaly-free global $U(1)$ symmetry, and an anomaly-free discrete symmetry.  As we can show that that   any operator allowed by the anomaly-free continuous global symmetries is allowed by the anomaly-free discrete symmetry, we will not discuss the discrete symmetry any further. This matter content is summarized in Tab.~\ref{matterUV}.

\begin{table}[htb!]
\centering 
\setlength{\tabcolsep}{8pt}
\renewcommand{\arraystretch}{1.2}
\begin{tabular}{c|ccccc}
  $$  & $Sp(6)$ & $SU(10)$ & $U(1)$\\

 \hline
 \\
$F$ & $\yng(1)$ & $\yng(1)$ &$ -\frac{4}{5}$\\
\\
$A$ & $\yng(2)$ & 1 & 1 \\
\end{tabular}
\caption{Matter content in UV (above confinement scale $\Lambda$)}
\label{matterUV}
\end{table}

Anomaly matching between the UV and the IR effective theory below the confinement scale, where only the flavor symmetry remains, will determine the low-energy degrees of freedom. We list some of the possible composite particles in this picture, following \cite{Csaki:1996eu}:
\begin{align}
T_k &=\Tr{A^k}, \quad k=2,3 \nonumber \\
M_k &= FA^kF, \quad k=0,1,2  \ .
\end{align}
The $T_k$ are singlets under $SU(10)$, and the $M_k$ are the antisymmetric part of the product of two fundamentals of $SU(10)$, which makes them $45$-plets. The composite matter quantum numbers are summarized in Tab.~\ref{matterIR}. 
\begin{table}[h!]
\centering 
\setlength{\tabcolsep}{8pt}
\renewcommand{\arraystretch}{1.2}
\begin{tabular}{c|ccccc}
  $$   & $SU(10)$&U(1) \\
 \hline
 \\
$T_k$ & 1 & $k$&  \\
\\
$M_{k}$ & $\yng(1,1)$ & $-\frac85+k$&\\
\end{tabular}
\caption{Matter content in IR (below confinement scale $\Lambda$)}
\label{matterIR}
\end{table}
The composite fermion $M_1=FAF$, which we will call $\Psi$, matches all of the $SU(10)\times U(1)$ global anomalies. If we assume that this theory has a phase which confines without chiral symmetry breaking, then $\Psi$ must be massless. We will also assume that   the composite scalar  $M_0=FF$   is anomalously light, and this meson we will call $\Phi$.   We assume the rest of the composite matter content has masses at the composite scale and thus  we  ignore them in the low-energy analysis.

In order to write down the low-energy effective theory we will need to consider the sources of explicit $SU(10)\times U(1)$ breaking. Part of the $SU(10)$ is weakly gauged under the Standard Model gauge group. Global baryon number symmetry is also embedded in the $SU(10)$. The fermions and their $Sp(6)\times SU(3)\times SU(2)_w\times U(1)_Y\times U(1)_B$ charges are listed in Tab.~\ref{preon_charge}, where the subscripts denote the component of $F$ that transforms in that representation under the Standard Model symmetries. 

\begin{table}[htb!]
\begin{tabular}{c|ccccc}
  $$  & Sp(6) &$SU(3)_c$ & $SU(2)_w$ & $U(1)_Y$&$U(1)_B$\\
 \hline
 \\
$F_3$ & $\yng(1)$&$\yng(1)$ & 1 &$\frac16$ &$\frac13$ \\
\\
$F_{\bar3}$ &$\yng(1)$& $\overline{\yng(1)}$ &1& $-\frac16 $  &$-\frac13$  \\
\\
$F_{2}$ & $\yng(1)$&1 & $\yng(1)$ & 0 &0\\
\\
$F_{+}$ &$\yng(1)$& 1 & 1 & +$\frac12$ &0\\
\\
$F_{-}$ &$\yng(1)$& 1 & 1 & $-\frac12$ &0\\
\\
$A$&$\yng(2)$&1&1&0&0\\
\\
\end{tabular}
\caption{Preon Charges}
\label{preon_charge}
\end{table}

The following $SU(10)\times U(1)$ global symmetry-breaking mass terms are consistent with the gauge symmetries and $ U(1)_B$ :

\be
{\cal L}\supset m_A AA + m_3  F_3   F_{\bar3} + m_2 F_2 F_2 + m_1 F_+ F_- +h.c. 
\ee

We will assume the $m_A$ and $m_3$ terms are small but non-negligible and the $m_2$ and $m_1$ terms are very small. The $SU(4)$ subgroup of the $SU(10)$ is thus mostly only explicitly broken by the weak gauge couplings. The $m_A$ and $m_3$ terms will lead to important spurions in the low-energy effective theory.

\section{\label{sec:comp} Effective Theory Below Compositeness Scale}
We will now analyze the effective dynamics of this model in the low-energy limit. The $m_3$ mass term in the UV for the $Q$ preons results in a spurion which explicitly breaks the $SU(6)$ part of the $SU(10)$ symmetry to $Sp(6)$ in the compositeness Lagrangian, leaving the $SU(4)$ invariant. The $m_A$ term does not break the $SU(10) $ but does break the $U(1)$, and will be treated as a spurion with $U(1)$ charge -2. 

Using the spurions to restore the $SU(10)\times U(1)$ symmetry, we write the non-kinetic terms in the compositeness Lagrangian that are invariant under this symmetry. In \eqn{lagrangian} we give the terms allowed through dimension 4 for the scalars and  up to  dimension 7 for  terms containing a fermion bilinear, where $M$ is the $Sp(6)$-preserving spurion whose $SU(6)$ part is given by
\be
M_{6\times6}=m_3\begin{pmatrix}
  0  & I_3 \\
    -I_3 &  0 \\ 
\end{pmatrix}  \ .
\ee

\begin{widetext}
\begin{eqnarray}
\mathcal{L} \ &&\supset (M_{10}^2 )\Tr \Phi^\dagger \Phi  -\Lambda^2\Tr M^\dagger\Phi   +\lambda_1 \Tr(\Phi^\dagger \Phi)^2+ \lambda_2 \Tr(\Phi^\dagger \Phi \Phi^\dagger \Phi ) +\frac{m_A}{\Lambda^2}\left[g_{\Psi}\Tr \Phi^{\dagger} \Psi \Phi^\dagger \Psi    \nonumber  \right. \\
&& \left.  +g_0 \Tr \Phi^{\dagger} \Psi    \Tr \Phi^{\dagger} \Psi +g_1 \Tr  M^\dagger \Psi     \Tr \Phi^\dagger \Psi  +g_2 \Tr M^{\dagger}  \Psi \Phi^\dagger \Psi   + g_3\Tr M^\dagger\Psi M^\dagger \Psi +g_4  \Tr M^\dagger \Psi\Tr M^\dagger \Psi  \right] \nonumber \\
&& + \lambda_3 \Tr(M^\dagger \Phi M^\dagger \Phi) +\lambda_4 \Tr(M^\dagger \Phi)\Tr( M^\dagger \Phi)+ \lambda_5 \Tr(M^\dagger \Phi \Phi^\dagger \Phi) +\lambda_6 \Tr(M^\dagger \Phi)\Tr( \Phi^\dagger \Phi)\nonumber \\
&& +\lambda_7 \Tr(M M^\dagger \Phi^\dagger \Phi  )+\lambda_8 \Tr(M \Phi^\dagger)\Tr( M^\dagger \Phi)+ \frac{g_7 m_A^*}{\Lambda^3}\epsilon^{abcdefghij}\Phi_{ab}\Phi_{cd}\Phi_{ef}\Psi_{gh}\Psi_{ij}+ h.c. 
\label{composite}
\end{eqnarray}\label{lagrangian}
\end{widetext}

Note that the term $M_{10}^2$ will depend on   $  m_A m_A^\dagger $ and on $ \Tr M^\dagger M$. This term is allowed by all the symmetries. If it is as large as the compositeness scale, then the power-counting of the effective theory will not be useful as either all the scalars would be too heavy to allow an effective field theoretic treatment or the chiral symmetry-breaking scale will be at the confinement scale and the top partners will be too heavy to belong in the effective theory. If, however,  $M_{10}$ is light compared with the compositeness scale, then an effective field theoretic treatment keeping all the scalars and fermions  can be quantitatively useful. Furthermore, this limit is technically natural as long as $M_{10}^2$ is no lighter than a loop correction to this mass squared.  Since the effective theory does not contain any couplings larger than ${\cal O}( 1)$, $M_{10}^2$ can be naturally be smaller than the compositeness scale by a factor of ${\cal O}(1/16\pi^2)$. In the following analysis, we will assume that $M_{10}^2$ is positive and that the chiral symmetry-breaking which leads to a pseudo-Goldstone boson Higgs is triggered by the explicit chiral symmetry-breaking, as would be the case if the theory were approximately supersymmetric. A analysis with a negative but small $M_{10}^2$ would lead to a similar spectrum. 

Under $SU(6) \times SU(4)$, the meson field decomposes as
\be
\Phi=\begin{pmatrix}
  \phi_X  & \phi_T \\
   -\phi_T^T &  \phi_Y \\ 
\end{pmatrix}  \ .
\ee
The spurion $M$ produces a tadpole that will lead to VEV for some components of $\phi_X$,
\be
\phi_0 =\frac{m_3 \Lambda^2}{2 M_{10}^2} \ ,
\ee
which must be smaller than $\Lambda$ in order for the effective compositeness Lagrangian to be useful.

Expanding around this VEV, the $\phi_X$ particles  have mass
\begin{eqnarray}
m_{\phi_X}^2 &=& 8\left(M_{10}^2+\lambda_6 m_3^2 +2m_3 \phi_0 (\lambda_4+3\lambda_5) \right. \nonumber \\
&&\left. \quad   -4\phi_0^2(3\lambda_1+\lambda_2)\right) \ .
\end{eqnarray}
The $\phi_T$ also get a mass from this VEV,
\begin{eqnarray}
m_{\phi_T}^2&=&4M_{10}^2+2m_3^2 \lambda_6+4 \phi_0 m_3(\lambda_4+6 \lambda_5)  \nonumber \\
&&  \quad -8 \phi_0^2(6 \lambda_1+\lambda_2) \ ,
\end{eqnarray}
as do the $\phi_Y$,
\be
m_{\phi_Y}^2=4M_{10}^2+24 \phi_0 m_3 \lambda_5-48 \phi_0^2\lambda_1 \ .
\ee
The $\phi_Y$ mass-squared can be taken to be negative so that  $\phi_Y$ obtains a VEV, spontaneously breaking $SU(4)$ to its $Sp(4)$ subgroup.   This will result in five pseudo-Nambu-Goldstone bosons, four of which  will play the role of our composite Higgs doublet.

\section{Spontaneous symmetry- breaking Effective Theory}

\subsection{$SU(4)/Sp(4)$ Higgs}
We can now analyze the spontaneous symmetry breaking in the Higgs sector. First, we expand $\phi_Y$ around the $Sp(4)$-preserving VEV 
\be
\Sigma_0=\begin{pmatrix}
  i\sigma_2  & 0 \\
    0 &  i\sigma_2 \\ 
\end{pmatrix}  \ ,
\ee
which also preserves the electroweak gauge group.

As in \cite{Katz:2005au}, the Nambu-Goldstone bosons of $SU(4)/Sp(4)$ symmetry breaking can be written as
\be
\langle \phi_Y\rangle=\Sigma =  f e^{i \Pi/f}\Sigma_0e^{i \Pi/f}
\qquad
\Pi=\begin{pmatrix}
  A  &  H \\
   H^\dagger & -A\\ 
\end{pmatrix}  \ ,
\ee
where
\be
H=\begin{pmatrix}
  h_0+i h_3  &  i h_2+ h_1 \\
   i h_2- h_1 & h_0-i h_3\\ 
\end{pmatrix} 
\qquad
A=\begin{pmatrix}
  \eta  &  0 \\
   0 & \eta\\ 
\end{pmatrix}  \ ,
\ee
and $f$ is the decay constant of the sigma model, which here is $f=\phi_0$.

In unitary gauge, where $h_1,h_2,h_3$ are eaten by the $W$ and $Z$ bosons, $\Sigma$ is given as:
\be
\Sigma=\begin{pmatrix}
  0  &  \cos\alpha & 0 & i\sin\alpha\\
   -\cos\alpha & 0 & -i\sin\alpha & 0\\ 
     0  &  i\sin\alpha & 0 & \cos\alpha\\
        - i \sin\alpha & 0 & -\cos\alpha & 0\\ 
\end{pmatrix}  \ ,
\ee
where $\alpha=\sqrt{h^2+\eta^2}/\sqrt{2}\phi_0$ and $h^2=\sum_i h_i^2$. 

We implement the symmetry breaking by setting $\phi_Y \to \phi_0 \Sigma$ in \eqn{composite}, the composite Lagrangian. We will see that the Higgs VEV
\be
\theta=\frac{\langle h \rangle}{\sqrt{2}\phi_0 } \ ,
\ee
will give mass to the top partners in the following section.

\subsection{Top Partner Embedding}
The symmetry breaking from the scalar sector will propagate to the fermion sector via the scalar-fermion couplings in \eqn{composite}, resulting in masses and Yukawa couplings for the composite fermions.

We can decompose the composite fermion $\Psi$ in terms of $SU(6)\times SU(4)$ as
\be
\Psi=\begin{pmatrix}
  X  & S \\
   -S^T &  Y \\ 
\end{pmatrix}  \ .
\ee
The top partners are in $S$, which as a $(\bf 6, \bf 4)$ of $SU(6)\times SU(4)$, decomposes into four components in terms of the global $SU(6)$ and the global $SU(4)$ which contains the  custodial $SU(2)_L \times SU(2)_R $. It is given by
\be
S=\begin{pmatrix}
  Q'  &  P' \\
   \bar{Q}' &  \bar{P}'\\ 
\end{pmatrix}  \ .
\ee
\eqn{pqrep} gives the decomposition in terms of $SU(3)_L \times SU(3)_R \times SU(2)_L \times SU(2)_R$ and   into $SU(3)_D \times SU(2)_L\times U(1)_Y$, where $U(1)_Y$ is a linear combination of the $T_3$ generator of $SU(2)_R$ and an $SU(10)$ generator. 
\begin{eqnarray}
Q' &=& (3, 1 ,2, 1) \to (3 ,2, 1/6) \nonumber \\
{P}' &=& (3, 1 ,1, 2) \to  (3 ,1, 2/3), (3 ,1, -1/3)  \nonumber \\
\bar{Q}'&=& (1, \bar{3} ,2, 1)  \to ( \bar{3} ,2, -1/6)\nonumber \\
\bar{P}' &=& (1, \bar{3} ,1, 2) \to  (\bar{3} ,1, 1/3), (\bar{3} ,1, -2/3) \ .
\label{pqrep}
\end{eqnarray}
We will gauge $SU(3)_D$ for QCD and $SU(2)_L$ for the weak group.   $Q'$ is the  top partner  which has the same gauge quantum numbers as the left-handed top quark and $\bar{P}'$ contains a particle with the same gauge quantum numbers as the left-handed anti-top, which we will call $\bar{T}'$.

\subsection{Light Composites}\label{composites}
We assume that below the compositeness scale, the low-energy effective  theory contains the Standard Model gauge bosons, three generations of quarks and leptons,   composite fermions which form a  45-plet of the $SU(10)$, and composite scalars which also form a  45-plet of the $SU(10)$. While the fermions are necessarily light  due to the approximate  $SU(10)\times U(1)$ global symmetry, there is no symmetry protecting the mass of the scalars. A description of such a symmetry-breaking transition may be made by introducing scalars, and if the critical value for the preon mass terms is low and the transition is second order then the scalars must be  light relative to the compositeness scale.
The preon charges in Table \ref{preon_charge} results in the charges for the composite fermions and bosons given in Tables~\ref{matter} and \ref{scalars}.

The fermion $U(1)$ charges are chosen so that we have an appropriately charged $Q$ and $T$. Note that because baryon number and lepton number are conserved, we have a   conserved discrete ``R-parity", $(-1)^{3B+L+2S}\equiv R_P$, which we list as well.

  \begin{table}[h!]
  \centering 
  \setlength{\tabcolsep}{4pt}
  \renewcommand{\arraystretch}{1.2}
  \begin{tabular}{c|cccccc}
  \hline
  Field  & $SU(3)_c$ & $SU(2)_L$ & $U(1)_{Y}$ & $U(1)_B$&$R_P$  \\
  \hline
  $L_i$ & 1 & \textbf{2} & -1/2 & 0 &1 \\
  $Q_{F,i}$ & \textbf{3} & \textbf{2} & 1/6 & 1/3 &1\\
  $\bar{U}_{F,i}$ & $\bar{\textbf{3}}$  & 1& -2/3 & -1/3&1 \\
  $\bar{D}_i$ & $\bar{\textbf{3}}$  & 1& 1/3 & -1/3 &1\\
  $\bar{E}_i$ & 1 & 1 &   1  & 0&1\\
\hline
 $Y$ & 1 & $\textbf{2}$  & 1/2 & 0 &-1\\
 $\tilde{Y}$ & 1 & $\textbf{2}$  & -1/2 & 0 &-1\\
  $\psi_1$ & 1 & 1 & 0 & 0&-1\\
   $\psi_2$ & 1 & 1  & 0 & 0&-1\\
   \hline
   $Q'$ & \textbf{3} & \textbf{2}  & $1/6$ & 1/3 &1\\
  ${T}'$ &  \textbf{3} & 1& $2/3$ & 1/3&1 \\
    ${B}'$ &  \textbf{3} & 1 & $-1/3$ & 1/3&1\\
   $\tilde{Q}'$ & $\bar{\textbf{3}}$ & \textbf{2} & $-1/6$ & -1/3 &1\\
 $\tilde{T}'$ & $\bar{\textbf{3}}$ & 1 & $-2/3$ & -1/3 &1 \\
  $ \tilde {B}'$ & $\bar{\textbf{3}}$ & 1 & $1/3$ & -1/3 &1 \\
     \hline
 $\tilde{D'}$ &$\bar{\textbf{3}}$  & 1 & $1/3$ & $2/3$&-1\\
 ${D}'$ &  \textbf{3} & 1 & $-1/3$ & $-2/3$&-1\\
  $\psi_3$ &  1 & 1 & 0 & 0 &-1\\
   $X$ &  \textbf{8} & 1  & 0  & 0&-1\\
  \end{tabular}
  \caption{Spin-1/2 field content of the low-energy effective theory. All fields are left-handed 2 component Weyl  spinors. }
  \label{matter}
  \end{table}
\begin{table}[h!]
  \centering 
  \setlength{\tabcolsep}{4pt}
  \renewcommand{\arraystretch}{1.2}
  \begin{tabular}{c|cccccc}
  \hline
  Field  & $SU(3)_c$ & $SU(2)_L$ & $U(1)_{Y}$ & $U(1)_B$&$R_p$ \\
\hline
 $h$ & 1 & $\textbf{2}$  & 1/2 & 0 & 1 \\
  $\eta$ & 1 & 1 & 0 & 0 & 1\\
   \hline
   $q' $ & \textbf{3} & \textbf{2}  & $1/6$ & 1/3  & -1\\
  ${t}' $ &  \textbf{3} & 1& $2/3$ & 1/3 & -1\\
    ${b}' $ &  \textbf{3} & 1 & $-1/3$ & 1/3 & -1\\
   $\tilde{q}' $ & $\bar{\textbf{3}}$ & \textbf{2} & $-1/6$ & -1/3 & -1 \\
 $\tilde{t}' $ & $\bar{\textbf{3}}$ & 1 & $-2/3$ & -1/3 & -1 \\
  $ \tilde {b}' $ & $\bar{\textbf{3}}$ & 1 & $1/3$ & -1/3 & -1 \\
     \hline
 $ \tilde{d}' $ &$\bar{\textbf{3}}$  & 1 & $1/3$ & $2/3$ & 1\\
 ${d}' $ &  \textbf{3} & 1 & $-1/3$ & $-2/3$ & 1\\
  $\rho$ &  1 & 1 & 0 & 0 & 1 \\
   $x$ &  \textbf{8} & 1  & 0  & 0 & 1\\
  \end{tabular}
  \caption{Spin-0 field content of low-energy effective theory. }
  \label{scalars}
  \end{table}
\subsection{Yukawas and Partial Compositeness}
We have obtained masses for the composite top partners via symmetry breaking, but it remains to propagate this to the fundamental top via partial compositeness \cite{Kaplan:1991dc}. This entails linearly coupling the composite quark and top partners  to the fundamental ones. We embed the fundamental degrees of freedom in an incomplete $SU(6) \times SU(4)$ multiplet:
\be
T_F=\begin{pmatrix}
  Q_F  & (\bar{T}_F \ 0) \\
  0 &  0 \\ 
\end{pmatrix}  \ ,
\ee
and couple it linearly to the quark doublet partner $\bar{Q}'$ and the anti-top partner $T'$:
\be
\mathcal{L} \supset g_5 \phi_0 \bar{T}_F T'+ g_6 \phi_0 Q_F \bar{Q}' \ .
\label{partial}
\ee
This gives the top partners a mass obtained from the charge-2/3 mass matrix $m_{2/3}$ given in Table~\ref{mass23}, where we have defined $g_t \equiv 2 m_A(g_2 m_3+2 g_{\psi}\phi_0)/\Lambda^2 $.

\begin{table}[htb]
  \centering 
  \setlength{\tabcolsep}{8pt}
  \renewcommand{\arraystretch}{1.2}
  \begin{tabular}{c|ccccc}
    & $\bar{Q}'$& $\bar{T}'$   & $\bar{T}_F$\\
  \hline
 $Q'$ & $g_t \phi_0 \cos{\theta}$ & $ig_t \phi_0 \sin{\theta}$    & 0 \\
   $T'$  & $ i g_t \phi_0  \sin{\theta}$  &$g_t \phi_0  \cos{\theta}$ & $g_5 \phi_0$  \\
 $Q_F$ & $g_6 \phi_0$ & 0  & 0\\ 
    \end{tabular}
  \caption{Charge-$2/3$ mass matrix}
  \label{mass23} 
  \end{table}
  
  Expanding to first order in $\theta$, we have
  \begin{eqnarray}
  \mathcal{L} \supset & \phi_0 \bar{Q}'(g_t Q'+g_6 Q_F)+\phi_0 T'( g_t  \bar{T}'+g_5  \bar{T}_F) \n
  &+i g_t \phi_0 T' \bar{Q}' \theta+i g_t \phi_0 Q' \bar{T}' \theta  \ .
  \label{topmass}
  \end{eqnarray}
We see that $\bar{Q}'$ pairs with the following linear combination of fields:
\be
\frac{ g_t Q' +g_6  Q_F}{\sqrt{ g_t^2+g_6^2 }} 
\ee
to gain a mass.

Similarly, $T'$ pairs with the combination:
\be
\frac{ g_t \bar{T}' +g_5  \bar{T}_F}{\sqrt{g_t^2+g_5^2}}
\ee
to become massive. 

The linear combinations of fields that are light are given by:
  \begin{eqnarray}
  T_L\equiv \frac{g_6  Q'- g_t Q_F}{\sqrt{ g_t^2+g_6^2 }} \n
    \bar{T}_R\equiv \frac{g_5  \tilde{T}'- g_t \bar{T}_F}{\sqrt{ g_t^2+g_5^2 }} \ .
   \end{eqnarray}
  These light quarks have mass term:
   \be
   \frac{\langle h \rangle}{\sqrt{2}} \epsilon_5 \epsilon_6 g_t T_L \bar{T}_R \ ,
\label{lighttopmass}
\ee
where we have defined the mixing angles
\begin{eqnarray}
\epsilon_5&& \equiv\frac{g_5}{\sqrt{g_t^2+g_5^2}} \n
\epsilon_6&&\equiv\frac{g_6}{\sqrt{g_t^2+g_6^2}} \ .
\end{eqnarray}

In contrast, the other quarks have masses $\phi_0 \sqrt{g_t^2+g_4^2}$ and $\phi_0 \sqrt{g_t^2+g_6^2}$, making them much heavier due to the fact that $\langle 
h \rangle \ll \phi_0$.

\subsection{Higgs Effective Potential}
As in \cite{Katz:2005au}, the contribution of the top to the Higgs effective potential (which dominates over the gauge boson contribution) is
\begin{align}
&-\sum_i 3 \frac{|m_i^2|^2}{16 \pi^2} \log|m_i^2|  ,
\end{align}
where $|m_i^2|$ are the eigenvalues of the charge-$2/3$ mass matrix $m_{2/3} m^\dagger_{2/3}$ obtained from \eqn{partial}.

The gauging of the EW symmetry does not break the $U(1)$ symmetry under which the $\eta$ singlet transforms, so it does not get a potential from gauge loops \cite{Kaplan:1991dc}. A massless $\eta$ would be ruled out by Kaon decays, so we must give it a small mass by adding another spurion 
\be
\phi_0^3 m_\eta \Tr(\Sigma_0^\dagger \Sigma) \ .
\ee
This term also contributes to the Higgs mass, so its contribution must be smaller than the electroweak breaking scale so as not to reintroduce fine-tuning.

\section{Phenomenology}

\subsection{Precision-Electroweak and Flavor Constraints}\label{PEW}

Any composite Higgs model must satisfy precision electroweak constraints. The most dangerous couplings are those of the bottom quark. At this point, the fundamental bottom quark is still massless. In order to give the bottom quark a mass we will need to couple the fundamental bottom to the composite quarks, for example with a coupling $g_B \bar{B}_F B $. In analogy to the generation of the top mass in \eqn{topmass}, this coupling will generate a mass for the bottom which is proportional to $g_B$. Since $g_B$ must be small in order to produce the observed small value of the bottom mass, the corrections to the $Z b \bar{b}$ coupling will be suppressed, as noted in \cite{Barbieri:2007bh,Vecchi:2013bja}. 

The $S$ is proportional to $\langle h \rangle^2/\phi_0^2$ \cite{Bellazzini:2014yua}, so it
does not receive too-large corrections as long the compositeness scale is larger than about 1 TeV. The custodial symmetry protects the $T$ parameter from large corrections \cite{Agashe:2006at}, but since the Yukawa couplings $g_5, g_6, g_B$ break the custodial symmetry we must consider how large these effects are. The $T$ parameter is modified by one-loop diagrams of the top partners and Standard Model fermions, which is parametrically of order
\be
\Delta T \sim \frac{m_t^2}{m_W^2} \epsilon_B^2 \frac{\langle h \rangle^2}{\phi_0^2} \log{\frac{g_t^2 \langle h \rangle^2 }{m_t^2}} 
\ee
in the limit $g_B \ll g_5,g_6$ \cite{Vecchi:2013bja}. We conclude that the modifications to the $T$ parameter are small because $\epsilon_B\ll 1$.

We next turn to flavor constraints. Because the top partners are weakly coupled to the Higgs via the partial compositeness mechanism, the scale of flavor violation can be decoupled from the compositeness scale. As shown in \cite{Cacciapaglia:2015dsa}, we can avoid large flavor-changing neutral currents by assuming that the fundamental Standard Model fermions are coupled to the $Sp(6)$ preons via four-fermion operators at some scale much larger than the compositeness scale $\Lambda$.

\subsection{Dark Matter Candidate}

As we discussed in section \ref{composites}, this theory has a conserved discrete symmetry related to lepton- and baryon-number symmetry analogous to an ``R-parity" for a supersymmetric theory, which we have termed $R_p$. This symmetry was previously discussed in the context of composite Higgs models in \cite{Katz:2003sn} and was termed ``dark matter parity". Neutral particles that are odd under this discrete symmetry are good dark matter candidates. From Table~\ref{matter} we see that the neutral, $R_p$-odd particles are the fermion singlets $\psi_a$. The lightest of these $R_p$-odd singlets could be the dark matter. 

\subsection{Particle Spectrum}\label{particlespectrum}

Since at the compositeness scale the complete $SU(10)$ fermion multiplet is massless and only gains mass from soft breaking terms, while the scalar has mass $M_{10}$ at the compositeness scale and gets additional soft corrections, we will assume all the composite fermions are lighter than the scalars except for the pseudo-Goldstones. Thus, to determine the novel phenomena of this model at the LHC we only need to consider the decays of the top partners to other fermions and to $h$ and $\eta$. 

\begin{table}[h!]
  \centering 
  \setlength{\tabcolsep}{4pt}
  \renewcommand{\arraystretch}{1.2}
  \begin{tabular}{c|cccccc}
   \hline
     Particle  & $SU(3)_c$ & $SU(2)_L$ & $U(1)_{Y}$ & $U(1)_B$&$R_p$ \\
     \hline
      $Y$ & 1 & $\textbf{2}$  & 1/2 & 0 &-1\\
 $\bar{Y}$ & 1 & $\textbf{2}$  & -1/2 & 0 &-1\\
  $\psi_1$ & 1 & 1 & 0 & 0&-1\\
   $\psi_2$ & 1 & 1  & 0 & 0&-1\\
      \hline
   $\begin{pmatrix}
  T_L  \\
   B_L\\ 
\end{pmatrix}  $ & \textbf{3} & \textbf{2}  & $1/6$ & 1/3 &1\\
  ${T}_R$ &  \textbf{3} & 1& $2/3$ & 1/3&1 \\
    ${B}_R$ &  \textbf{3} & 1 & $-1/3$ & 1/3&1\\
    $\begin{pmatrix}
  T'_L  \\
   B'_L\\ 
\end{pmatrix}  $ & \textbf{3} & \textbf{2}  & $1/6$ & 1/3 &1\\
  ${T}'_R$ &  \textbf{3} & 1& $2/3$ & 1/3&1 \\
    ${B}'_R$ &  \textbf{3} & 1 & $-1/3$ & 1/3&1\\
   $\begin{pmatrix}
 \bar {T}''_L  \\
  \bar {B}''_L\\ 
\end{pmatrix}  $ & $\bar{\textbf{3}}$ & \textbf{2} & $-1/6$ & -1/3 &1\\
 $\bar{T}''_R$ & $\bar{\textbf{3}}$ & 1 & $-2/3$ & -1/3 &1 \\
  $\bar{B}''_R$ & $\bar{\textbf{3}}$ & 1 & $1/3$ & -1/3 &1 \\
     \hline
 $\bar{D}'$ &$\bar{\textbf{3}}$  & 1 & $1/3$ & $2/3$&-1\\
 ${D}'$ &  \textbf{3} & 1 & $-1/3$ & $-2/3$&-1\\
  $\psi_3$ &  1 & 1 & 0 & 0 &-1\\
   $X$ &  \textbf{8} & 1  & 0  & 0&-1\\
  \end{tabular}
  \caption{Fermionic Particle Content (Mass Basis)}
  \label{particles}
  \end{table}
  
As summarized in Table \ref{particles}, our theory contains the light top quark $T$ and two heavy quarks $T'$ and $\bar{T}''$, as well as the bottom $B$ and its partners $B'$ and $\bar{B}''$, just like the $SU(4)/Sp(4)$ intermediate Higgs model \cite{Katz:2005au}, but it also contains the uncolored weak doublets $Y$ and $\bar{Y}$, the color triplets with exotic baryon number $D'$ and $\bar{D}'$, a color octet $X$, and three neutral singlets $\psi_1, \psi_2, \psi_3$. In addition, we have the $SU(4)/Sp(4)$ Higgs doublet $h$ and the additional scalar singlet $\eta$. 

The phenomenology of this model will depend on the particle spectrum. Masses for the composite fermions are generated by the spurion $M$ and the VEV of $\phi$:
\be
\phi_0=\frac{m_3 \Lambda^2}{2 M_{10}^2} < \Lambda \ .
\label{phi0def}
\ee
From this expression we see that  $m_3/\phi_0< \Lambda^2/2 M_{10}^2$, and so $m_3$ cannot be parametrically larger than $\phi_0$ unless $M_{10}$ is larger than $\Lambda$, which would break our perturbative expansion. Thus from ~\eqn{composite}, we expect that the new fermions will all have mass of order $m_A  \phi_0^2 /\Lambda^2$. In contrast, the scalars have typical mass-squared of order $\phi_0^2$, so since we must have $m_A \ll \Lambda$ and $\phi_0<\Lambda$ in order for the effective compositeness Lagrangian to be consistent, this means that the non-pseudo-Goldstone scalars are generically heavier than the fermions. We can therefore neglect them in our analysis of the low-energy phenomenology.

The partial compositeness described in section \ref{PEW} mixes $T', \bar{T}',T'', \bar{T}'',Q, \bar{Q}$ with the fundamental quarks $T_F, Q_F$, resulting in the physical quarks $T_L$ and $\bar{T}_R$ as well as two heavier species of quarks. This mechanism results in a top mass of order 
\be
m_T \sim \epsilon_5 \epsilon_6 \frac{m_A \phi_0}{\Lambda^2}\langle h \rangle \ ,
\ee
which must be matched to the observed value.  The fermions which are not top partners have masses of order
\be
m_{\rm exotic} \sim g_*\frac{m_A}{\Lambda^2}\phi_0^2 \ ,
\ee
where $g_*$ is a linear combination of the coupling parameters ${g_{\psi}, g_0, g_1, g_2, g_3, g_4}$. 

In addition to the top and bottom partners, the particles most relevant for phenomenology are the Higgs and the singlet $\eta$, where $m_\eta < m_h$. The ratio of the exotic fermion masses to the Higgs sector masses is
\be
\frac{m_{\rm exotic}}{\langle h \rangle}  \sim g_* \frac{m_A}{\Lambda} \frac{\phi_0}{\Lambda}\frac{\phi_0}{\langle h \rangle}  \ ,
\ee
where $\phi_0/\langle h \rangle >1$, $\phi_0/\Lambda<1$, and  $m_A /\Lambda \ll 1$. Note that it is possible for some of the exotic fermions to be lighter than $h$ and $\eta$, but since exotic decays of the Higgs are strongly constrained by the LHC, we will not consider this case.

The masses of the exotic fermions are given in terms of the couplings in the Lagrangian as:
\begin{align}
m_Y&=m_{\bar{Y}}=\frac{2 g_\psi m_A \phi_0^2}{\Lambda^2} \nonumber \\
m_{\psi_1}&=m_Y \nonumber \\
m_{\psi_2}&=m_Y+\frac{8 g_0 m_A \phi_0^2}{\Lambda^2} \nonumber \\
m_X&=\frac{2  m_A }{\Lambda^2}\left (g_3m_3^2+g_2\phi_0 m_3+g_\psi\phi_0^2 \right) \nonumber \\ 
m_D&=m_{\bar{D}}=2m_X  \nonumber \\
m_{\psi_3}&=m_X+\frac{12  m_A }{\Lambda^2}\left (g_4m_3^2+g_1\phi_0 m_3+  g_0 \phi_0^2 \right ) , \label{mpsi}
\end{align}
and the heavy top partners have masses 
\begin{align}
m_{T'}= \phi_0\sqrt{g_t^2+g_5^2}\nonumber \\
m_{T''}= \phi_0\sqrt{g_t^2+g_6^2} ,
\label{heavytopmass}
\end{align}
where $g_t=2m_A(g_2 m_3+2 g_\psi \phi_0)/\Lambda^2$. 

\subsubsection{Spectrum for $m_3 \ll \phi_0$}

For simplicity we will first study the regime where $m_3/\phi_0 \ll 1$ so that at leading order we can neglect the terms proportional to $m_3$ in the masses. We see that in this limit, $m_X \approx  m_Y$ and $m_{\psi_3} \approx  m_Y+12 g_0 \phi_0^2 m_A /\Lambda^2  $. Thus $Y,\bar{Y}, \psi_1$, and $X$ have mass $m_Y$ while $D' \bar{D}'$ are heavier, with a mass of approximately $2 m_Y$. The remaining fermions $\psi_2$ and $\psi_3$ 
are either lighter or heavier than $m_Y$, depending on the sign of $g_0$. If $g_0<0$, then $m_{\psi_3}<m_{\psi_2}<m_Y$, and $m_{\psi_3}$ is the dark matter candidate. The six parameters $g_i$ have collapsed in this limit to two independent parameters, $m_Y$ and $m_{\psi_3}$ because the terms involving $g_1,g_2,g_3,g_4$ are negligible.

Finally, we need to consider how the heavy top partners fit into the mass hierarchy. Since $\phi_0/m_3 \gg 1$, we find that 
\be
g_t \approx \frac{4 g_\psi m_A \phi_0}{\Lambda^2} .
\ee
Thus, from \eqn{heavytopmass} we see that in the limit of no mixing with the fundamental quarks ($g_5=g_6=0$), $T'$ and $T''$ both have mass $2 m_Y$, which is equal to that of the exotic triplets $D',\bar{D}'$. When $g_4$ and $g_5$ are nonzero, $Q'$ and $Q''$ become heavier than $D',\bar{D}'$ because $\sqrt{g_t^2+g_i^2}>g_t$. This makes them the heaviest of the exotic fermions.

The masses of heavy top partners are bounded from above by the inverse of the fine-tuning parameter $\langle h \rangle /\phi_0$. In the absence of fine-tuning they should be  lighter than about $800$ GeV, and they are bounded from below by LHC searches. Since the $X$ particle is lighter than the top partners in this case, there are strong bounds from light gluino searches on this regime of the model, which we will discuss in section \ref{sec:pheno}.

\subsubsection{Spectrum for $m_3 \sim \phi_0$}
The other regime of our model that is allowed by perturbativity is where $m_3$ and $\phi_0$ are comparable. To understand what this relation implies about the masses in the theory, we can rewrite $M_{10}$ and $m_3$ in terms of dimensionless parameters, $M_{10}=\epsilon \Lambda$, $m_3=\beta \Lambda$. Then the constraints imposed by requiring  our theory to be perturbative are $\phi_0<\Lambda$, $M_{10}<\Lambda$, and $m_3 < \Lambda$. In terms of the dimensionless parameters, this is $\epsilon<1$, $\beta<1$,  $\frac{\beta}{2 \epsilon^2} < 1$. Since $m_3/\phi_0 =2 \epsilon^2$, 
we can take $m_3 \sim \phi_0$ as long as $\epsilon$ is a fraction of order one and $\beta \ll 2\epsilon^2$. This corresponds to $M_{10}$ close to (but less than) the compositeness scale, and $m_3$ much lighter.

In the limit $m_3 \sim \phi_0$, the masses $m_Y$, $m_{\psi_2}$, $m_X$, and $m_{\psi_3}$ are independent parameters. In addition to these mass parameters, we have the top sector masses, $m_T$, $m_{T'}$, and $m_{T''}$, which are also independent parameters because they depend on $g_2$, $g_{\psi}$, $g_5$, and $g_6$ in the Lagrangian. Thus we can trade seven of the eight original dimensionless parameters $g_i$ in the Lagrangian for the seven masses. Only the pair $(g_1,g_4)$ remains degenerate in \eqn{mpsi}.

\subsection{\label{sec:pheno} LHC Phenomenology}

The strongest phenomenological constraints on our scenario comes from the LHC. The phenomenology of the Higgs plus singlet arising from the $SU(4)/Sp(4)$ coset structure was previously studied in \cite{Gripaios:2009pe,Arbey:2015exa}, but these analyses did not include top partners. Vector-like quark partners coupling to the top that are singlets or doublets under the  custodial $SU(2) \times SU(2)$ have been constrained to have masses greater than around 800 GeV after the first run of the LHC \cite{Backovic:2014uma}, but additional  decay modes of the top partners will open new channels for searches at the LHC. In addition, the possibility of producing $\eta$ will affect the decays of these exotic vector-like quarks.

The standard channels to look for new vector-like quarks are through the decays $T \to b W$, $T \to t Z$, and $T \to t h$. Because our model has a light pseudoscalar $\eta$, we have the additional processes $T \to t \eta$ and $B \to b \eta$. These decays were studied in \cite{Bizot:2018tds} with the conclusion that the branching ratios to these exotic decays can be large and  of the order of the Standard Model decays. Since limits on the top partner mass depend on the branching ratios to Standard Model particles, large branching ratios to $\eta$ can significantly affect these limits. In addition to $\eta$, the exotic fermions $D',\bar{D}'$, $X$, and the $\psi_a$ will affect searches for the top partners. 

 \begin{table}[h!]
    \centering 
    \setlength{\tabcolsep}{8pt}
    \renewcommand{\arraystretch}{1.2}
    \begin{tabular}{|c|c|}
  \hline
dim 4  & $h \bar{Y} \psi_a$\\
&  $h \bar{Y} X $ \\
&$\eta \psi_a \psi_b$ \\
\hline
dim 5&  $ \bar{Y} \psi_a h \eta$\\
&  $ \bar{Y} X h \eta$\\
& $ \psi_a \psi_b \eta \eta $\\
\hline
dim 6&  $ \psi_a D T_L^{i} B_L^j $\\
& $ \psi_a D T_R^{i} B_R^j $\\
& $\psi_a X T_L^{i} \bar{T}_L^j$\\
& $\psi_a X B_L^{i} \bar{B}_L^j$\\
& $\psi_a X T_R^{i} \bar{T}_R^j$\\
& $\psi_a X B_R^{i} \bar{B}_R^j$\\
& $ \psi_a \psi_b X X $\\
&$ \psi_a \psi_b D \bar{D} $\\
&$ \psi_a \psi_b Y \bar{Y} $\\
&$ \psi_a \psi_b \psi_c \psi_d $\\
&$\psi_a \psi_b T_L^{i} \bar{T}_L^j $\\
  &       $\psi_a \psi_b B_L^{i} \bar{B}_L^j $\\
  &      $\psi_a \psi_b T_R^{i} \bar{T}_R^j $\\
  &       $\psi_a \psi_b B_R^{i} \bar{B}_R^j $\\
\hline 
\end{tabular}
\caption{Leading interactions producing decay of the exotic fermions. Here, the superscript $i$ refers to unprimed, primed, or double primed for the three generations of top and bottom quarks, and the subscript $a$ ranges over ${1,2,3}$ for the three neutral fermions.}
    \label{decay}
    \end{table}

To see this, consider Table~\ref{decay}, which lists the leading interactions involving the decay of the fermions in our model up to dimension 6 (the full list of interactions including processes other than decays is given in Table~\ref{inter45} and Table~\ref{inter} in Appendix~\ref{appint}). In addition to these, the dimension-4 Higgs sector interactions  $h\bar{h} \eta \eta,h\bar{h} h\bar{h}, \eta \eta \eta \eta $ are relevant for LHC phenomenology.

\subsubsection{LHC constraints for $m_3 \ll \phi_0$}

The decay processes depend on the mass hierarchy between the exotic fermions $Y,X,\psi_a,D',\bar{D}'$. We will first explore the case $m_3 \ll \phi_0$. From the discussion in section~\ref{particlespectrum}, we know that if  $g_0<0$ then the mass spectrum is $m_{T''}\sim m_{T'} > m_D>m_Y \sim m_X \sim m_{\psi_1} >m_{\psi_2} >m_{\psi_3}$. Any colored particle will be produced from gluons at the LHC, so we expect the color octet $X$ and the $D',\bar{D}'$ to be produced as well as the top partners. 

Since decays of $X$ to $Y$ are not kinematically allowed, the dominant decay process of $X$ is to $t\bar{t} \psi_a$ at dimension 6. This means that $X$ acts like a gluino in the case of a heavy  stop, and so it decays to $t, \bar{t}$ and a neutralino LSP, which in our case is the dark matter candidate $\psi_3$ \cite{Aad:2013wta}.

The main signatures of this model is large missing energy plus multi-jets, which is similar to searches for natural SUSY \cite{Evans:2013jna,Buckley:2016kvr}. These searches constrain the gluinos to be heavier than around $1.2$ TeV. This is larger than the upper bound on the top partner mass, so this regime is essentially ruled out. One consideration that tends to weaken the constraint from gluino searches is that if the parameter $g_0$ is small, then there is a small mass difference between the ``gluino" and the ``neutralino", which leads to less energetic jets and missing transverse energy, which can weaken the limits on the ``gluino" mass \cite{Tanabashi:2018oca}. However, we do not expect this weakening to accommodate the required mass of less than $800$ GeV for the $X$. Furthermore, the additional degrees of freedom of the Weyl fermion $X$ with respect to the scalar gluino serve to strengthen the limits. Thus, the $m_3 \ll \phi_0$ regime of our model is not  viable in a natural mass regime.

\subsubsection{LHC constraints for $m_3 \sim \phi_0$}
We turn to the regime where $m_3$ is of the same order as $\phi_0$. We have several experimental constraints to bound the seven free masses that determine the theory in the case $m_3 \sim \phi_0$: $m_T$ is fixed by its observed value, and $m_{T'}$ and $m_{T''}$ are constrained by naturalness to be less than about $800$ GeV. Since in this case $m_X$ is a free parameter, it can be larger than the mass of the heavy top partners, which would alleviate the strong bounds from light gluinos. In addition, since the singlets and the $Y$ could be lighter than the $X$, there can be additional decay modes to these particles plus the Higgs starting at dimension 4, which we can see from Table~\ref{decay}. Since $m_D$ is fixed by this model to have mass $2 m_X$, $D', \bar{D}'$ will be heavier than the top partners and can decay to them via dimension-6 operators.

The singlet masses $m_{\psi_2}$ and $m_{\psi_3}$ are not strongly constrained by the LHC, and different hierarchies of these masses can open up additional decay modes for the top partners in Table~\ref{decay}. Note that the other singlet $m_{\psi_1}$ is fixed to have the same mass as $Y$. If either of $\psi_2$ or $\psi_3$ is the lightest of the $R_p$-odd fermions in this model, it can be a dark matter candidate.

This regime of our model has a particle and interaction content that combines aspects of SUSY  with that of composite Higgs models. The SUSY-like features, include gluino-, higgsino-, and neutralino-like particles,and a conserved dark matter parity. Composite-like features include two vector-like heavy top partners and an additional pseudoscalar in the Higgs sector.

\section{\label{sec:discussion}Discussion}

The discovery of the Higgs boson at the LHC in the absence of any other new particles has left us few clues to understand its unnaturally small mass. One possibility is that the Higgs boson is light because it is a composite particle of some underlying strongly-coupled gauge theory. In the strongly-coupled regime perturbative calculations cannot be made but lattice calculations can shed light on its spectrum. The low-energy spectrum of such a theory is expected to have complex phenomenology due to the large number of possible composite particles of the fundamental degrees of freedom, in contrast to the simplest composite Higgs models which do not address the UV completion of the theory. Thus we expect realistic models of a composite Higgs to have richer phenomenology than minimal models.

We find that at low energies, this model indeed has several new fermions: a color octet, an electroweak doublet, three scalars, two vector-like quark partners, and a pair of color triplets with exotic baryon number. This theory exhibits phenomenology that combines aspects of SUSY theories with aspects of composite Higgs theories, and can be studied on the lattice to see whether our expectations about the low-energy spectrum are borne out. A recent lattice study of partial compositeness in a $SU(4)$ gauge theory
suggests that the spectrum of our model can indeed be probed \cite{Ayyar:2018glg}. 
They found that the $SU(4)$ model motivated by \cite{Ferretti:2013kya,Ferretti:2014qta} is not able to produce a realistic top mass and attribute this to the fact that the theory did not have the required near-conformal dynamics to generate small composite fermion masses. In our model, the small fermion mass is due to the theory being confining without breaking the chiral symmetries and having matter content which is near a second-order phase transition so that the chiral symmetry can be spontaneously broken. These assumptions can be checked on the lattice since our fundamental matter content is vector-like. Thus, our model is a good candidate to produce realistic top couplings in a lattice study like \cite{Ayyar:2018glg}. 

Since this model is the minimal one that can serve as a UV completion of the composite Higgs with top partners and which has a fundamental matter content that can be simulated on the lattice, models with larger symmetry groups will likely provide novel phenomenology accessible to the LHC. A promising avenue for future work is to explore such non-minimal ``lattice-friendly" models.

\begin{acknowledgments}
A.N. is funded in part by the DOE under grant DE-SC0011637, and by the Kenneth K. Young Chair.  H.G. acknowledges partial support from the Danish National Research Foundation grant DNRF:90, the Fulbright commission and the Ole R\o mer foundation.  
\end{acknowledgments}

 \appendix
 \section{Interactions of Composite Fermions up to Dimension 6} \label{appint}
Here we list all the allowed operators including the composite fermions up to dimension 6. The subset of these that are relevant for decays are given in Table~\ref{decay}. Table~\ref{inter45} lists all dimension-4 and dimension-5 operators involving composite fermions, and Table~\ref{inter} lists the dimension-6 operators.
\begin{table}[htb]
    \centering 
    \setlength{\tabcolsep}{8pt}
    \renewcommand{\arraystretch}{1.2}
    \begin{tabular}{|c|c|}
    \hline
Dim 4 & Dim 5 \\
  \hline
  $h T_R^i \bar{T}_L^j$ &    $ D \bar{D} \eta \eta$  \\
 $h B_L^i \bar{B}_R^j$ &      $ Y \bar{Y} \eta \eta $  \\
  $h \bar{Y} \psi_a$   &    $ \psi_a \psi_b \eta \eta $ \\
   $h \bar{Y} X$      &    $X X \eta \eta $ \\
    $\eta Y \bar{Y}$ &     $\bar{Y} \bar{Y} h h $      \\
 $\eta \psi_a \psi_b$  &     $T_L^i \bar{T}_L^j \eta \eta $ \\
$\eta X X$   & $T_R^i \bar{T}_R^j \eta \eta $  \\
 $\eta D \bar{D}$  & $B_R^i \bar{B}_R^j \eta \eta $  \\
  $\eta B_L^i \bar{B}_L^j$  &$B_L^i \bar{B}_L^j \eta \eta $ \\
 $\eta B_R^i \bar{B}_R^j$ & $  B_R^i \bar{T}_R^j h h $  \\
   $\eta T_L^i \bar{T}_L^j$&$T_R^i \bar{T}_L^j h\eta$ \\
  $\eta T_R^i \bar{T}_R^j$   &$ B_L^i \bar{B}_R^j h \eta$ \\
  &$ \bar{Y} \psi_a h \eta$ \\
 & $ \bar{Y} X h \eta$    \\
 \hline
      \end{tabular}
    \caption{Dimension-4 and dimension-5 operators of fermions for LHC phenomenology. Here, the superscript $i$ refers to unprimed, primed, or double primed for the three generations of top and bottom quarks, and the subscript $a$ ranges over ${1,2,3}$ for the three neutral fermions.}
    \label{inter45}
    \end{table}

    \begin{table}[htb]
    \centering 
    \setlength{\tabcolsep}{3pt}
    \renewcommand{\arraystretch}{1.2}
    \begin{tabular}{|c|c|c|c|}
    \hline
     $D \bar{D} X X$&      $\psi_a \psi_b T_L^{i} \bar{T}_L^j $ &     $ Y\bar{Y} T_L^{i} \bar{T}_L^j $ &          $T_L^{i} \bar{T}_R^j T_R^k \bar{T}_L^m $ \\
    $Y \bar{Y} X X$ &       $\psi_a \psi_b B_L^{i} \bar{B}_L^j $ &      $ Y\bar{Y} B_L^{i} \bar{B}_L^j $ &          $ T_L^{i} \bar{T}_L^j T_L^k \bar{T}_L^m $ \\ 
    $ \psi_a \psi_b X X $ &      $\psi_a \psi_b T_R^{i} \bar{T}_R^j $ &   $ Y\bar{Y} T_R^{i} \bar{T}_R^j $ &                  $ T_R^{i} \bar{T}_R^j T_R^k \bar{T}_R^m $ \\ 
        $Y \bar{Y} D \bar{D}$&       $\psi_a \psi_b B_R^{i} \bar{B}_R^j $ &        $ Y\bar{Y} B_R^{i} \bar{B}_R^j $   &  $ B_L^{i} \bar{B}_L^j B_R^k \bar{B}_R^m $ \\
  $\psi_a \psi_b D \bar{D}$ &  $ \psi_a D T_L^{i} B_L^j $ & $ \bar{Y}\bar{Y} T_R^{i} \bar{B}_R^j $ &      $ B_L^{i} \bar{B}_L^j B_L^k \bar{B}_L^m $\\
  $XX X X $ &        $ \psi_a D T_R^{i} B_R^j $ &     $YY B_R^{i} \bar{T}_R^j$ &      $ B_R^{i} \bar{B}_R^j B_R^k \bar{B}_R^m$  \\ 
  $D \bar{D}  D \bar{D}$ &   $D \bar{D} T_L^i \bar{T}_L^j$   &    $\bar{Y}{Y} T_R^{i} \bar{B}_R^j$ &    $ T_L^{i} \bar{T}_L^j B_R^k \bar{B}_R^m $\\
  $Y \bar{Y}  \psi_a \psi_b $ &   $D \bar{D} B_R^i \bar{B}_R^j $   &    $XX T_L^{i} \bar{T}_L^j$ &     $T_R^{i}  \bar{T}_R^j \bar{B}_L^k B_L^m$\\
  $ \psi_a \psi_b  \psi_c \psi_d  $&  $D \bar{D}  T_R^i \bar{T}_R^j $    &    $XX B_L^{i} \bar{B}_L^j$ &    $T_R^{i} \bar{T}_R^j B_R^k \bar{B}_R^m $\\
  $ Y\bar{Y} Y  \bar{Y}$ &   $D \bar{D} B_L^i \bar{B}_L^j $    &   $XX T_R^{i} \bar{T}_R^j$ &   $ T_L^{i} \bar{T}_L^j B_L^k \bar{B}_L^m $\\
  $\psi_a X B_L^{i} \bar{B}_L^j$ &  $\psi_a X B_R^{i} \bar{B}_R^j$ &  $XX B_R^{i} \bar{B}_R^j$ & $T_L^{i} \bar{T}_R^j B_L^k \bar{B}_R^m $\\ 
   $\psi_a X T_R^{i} \bar{T}_R^j$&  $\psi_a X T_L^{i} \bar{T}_L^j$  &   &   $ \bar{T}_L^{i} T_R^j  \bar{B}_L^k B_R^m$\\
\hline
   \end{tabular}
    \caption{Dimension-6 operators of fermions for LHC phenomenology. Here, the superscript $i$ refers to unprimed, primed, or double primed for the three generations of top and bottom quarks, and the subscript $a$ ranges over ${1,2,3}$ for the three neutral fermions.}
    \label{inter}
    \end{table}

\bibliographystyle{JHEP}

\bibliography{lattice-composite_bib}

\providecommand{\href}[2]{#2}\begingroup\raggedright\begin{thebibliography}{10}

\bibitem{Katz:2003sn}
E.~Katz, J.-y. Lee, A.~E. Nelson and D.~G.~E. Walker, \emph{{A Composite little
  Higgs model}},
  \href{https://doi.org/10.1088/1126-6708/2005/10/088}{\emph{JHEP} {\bfseries
  10} (2005) 088} [\href{https://arxiv.org/abs/hep-ph/0312287}{{\ttfamily
  hep-ph/0312287}}].

\bibitem{Georgi:1984af}
H.~Georgi and D.~B. Kaplan, \emph{{Composite Higgs and Custodial SU(2)}},
  \href{https://doi.org/10.1016/0370-2693(84)90341-1}{\emph{Phys. Lett.}
  {\bfseries 145B} (1984) 216}.

\bibitem{ArkaniHamed:2002qy}
N.~Arkani-Hamed, A.~G. Cohen, E.~Katz and A.~E. Nelson, \emph{{The Littlest
  Higgs}}, \href{https://doi.org/10.1088/1126-6708/2002/07/034}{\emph{JHEP}
  {\bfseries 07} (2002) 034}
  [\href{https://arxiv.org/abs/hep-ph/0206021}{{\ttfamily hep-ph/0206021}}].

\bibitem{Cacciapaglia:2015vrx}
G.~Cacciapaglia and A.~Parolini, \emph{{Light ’t Hooft top partners}},
  \href{https://doi.org/10.1103/PhysRevD.93.071701}{\emph{Phys. Rev.}
  {\bfseries D93} (2016) 071701}
  [\href{https://arxiv.org/abs/1511.05163}{{\ttfamily 1511.05163}}].

\bibitem{Brower:2015owo}
R.~C. Brower, A.~Hasenfratz, C.~Rebbi, E.~Weinberg and O.~Witzel,
  \emph{{Composite Higgs model at a conformal fixed point}},
  \href{https://doi.org/10.1103/PhysRevD.93.075028}{\emph{Phys. Rev.}
  {\bfseries D93} (2016) 075028}
  [\href{https://arxiv.org/abs/1512.02576}{{\ttfamily 1512.02576}}].

\bibitem{Kaplan:1983sm}
D.~B. Kaplan, H.~Georgi and S.~Dimopoulos, \emph{{Composite Higgs Scalars}},
  \href{https://doi.org/10.1016/0370-2693(84)91178-X}{\emph{Phys. Lett.}
  {\bfseries 136B} (1984) 187}.

\bibitem{Katz:2005au}
E.~Katz, A.~E. Nelson and D.~G.~E. Walker, \emph{{The Intermediate Higgs}},
  \href{https://doi.org/10.1088/1126-6708/2005/08/074}{\emph{JHEP} {\bfseries
  08} (2005) 074} [\href{https://arxiv.org/abs/hep-ph/0504252}{{\ttfamily
  hep-ph/0504252}}].

\bibitem{Csaki:1996zb}
C.~Csaki, M.~Schmaltz and W.~Skiba, \emph{{Confinement in N=1 SUSY gauge
  theories and model building tools}},
  \href{https://doi.org/10.1103/PhysRevD.55.7840}{\emph{Phys. Rev.} {\bfseries
  D55} (1997) 7840} [\href{https://arxiv.org/abs/hep-th/9612207}{{\ttfamily
  hep-th/9612207}}].

\bibitem{Vafa:1983tf}
C.~Vafa and E.~Witten, \emph{{Restrictions on Symmetry Breaking in Vector-Like
  Gauge Theories}},
  \href{https://doi.org/10.1016/0550-3213(84)90230-X}{\emph{Nucl. Phys.}
  {\bfseries B234} (1984) 173}.

\bibitem{Preskill:1981sr}
J.~Preskill and S.~Weinberg, \emph{{'DECOUPLING' CONSTRAINTS ON MASSLESS
  COMPOSITE PARTICLES}},
  \href{https://doi.org/10.1103/PhysRevD.24.1059}{\emph{Phys. Rev.} {\bfseries
  D24} (1981) 1059}.

\bibitem{Barnard:2013zea}
J.~Barnard, T.~Gherghetta and T.~S. Ray, \emph{{UV descriptions of composite
  Higgs models without elementary scalars}},
  \href{https://doi.org/10.1007/JHEP02(2014)002}{\emph{JHEP} {\bfseries 02}
  (2014) 002} [\href{https://arxiv.org/abs/1311.6562}{{\ttfamily 1311.6562}}].

\bibitem{Ferretti:2013kya}
G.~Ferretti and D.~Karateev, \emph{{Fermionic UV completions of Composite Higgs
  models}}, \href{https://doi.org/10.1007/JHEP03(2014)077}{\emph{JHEP}
  {\bfseries 03} (2014) 077} [\href{https://arxiv.org/abs/1312.5330}{{\ttfamily
  1312.5330}}].

\bibitem{Ferretti:2014qta}
G.~Ferretti, \emph{{UV Completions of Partial Compositeness: The Case for a
  SU(4) Gauge Group}},
  \href{https://doi.org/10.1007/JHEP06(2014)142}{\emph{JHEP} {\bfseries 06}
  (2014) 142} [\href{https://arxiv.org/abs/1404.7137}{{\ttfamily 1404.7137}}].

\bibitem{Cacciapaglia:2014uja}
G.~Cacciapaglia and F.~Sannino, \emph{{Fundamental Composite (Goldstone) Higgs
  Dynamics}}, \href{https://doi.org/10.1007/JHEP04(2014)111}{\emph{JHEP}
  {\bfseries 04} (2014) 111} [\href{https://arxiv.org/abs/1402.0233}{{\ttfamily
  1402.0233}}].

\bibitem{Cacciapaglia:2015eqa}
G.~Cacciapaglia, H.~Cai, A.~Deandrea, T.~Flacke, S.~J. Lee and A.~Parolini,
  \emph{{Composite scalars at the LHC: the Higgs, the Sextet and the Octet}},
  \href{https://doi.org/10.1007/JHEP11(2015)201}{\emph{JHEP} {\bfseries 11}
  (2015) 201} [\href{https://arxiv.org/abs/1507.02283}{{\ttfamily
  1507.02283}}].

\bibitem{Vecchi:2015fma}
L.~Vecchi, \emph{{A dangerous irrelevant UV-completion of the composite
  Higgs}}, \href{https://doi.org/10.1007/JHEP02(2017)094}{\emph{JHEP}
  {\bfseries 02} (2017) 094}
  [\href{https://arxiv.org/abs/1506.00623}{{\ttfamily 1506.00623}}].

\bibitem{Sannino:2016sfx}
F.~Sannino, A.~Strumia, A.~Tesi and E.~Vigiani, \emph{{Fundamental partial
  compositeness}}, \href{https://doi.org/10.1007/JHEP11(2016)029}{\emph{JHEP}
  {\bfseries 11} (2016) 029}
  [\href{https://arxiv.org/abs/1607.01659}{{\ttfamily 1607.01659}}].

\bibitem{Csaki:1996eu}
C.~Csaki, W.~Skiba and M.~Schmaltz, \emph{{Exact results and duality for SP(2N)
  SUSY gauge theories with an antisymmetric tensor}},
  \href{https://doi.org/10.1016/S0550-3213(96)00709-2}{\emph{Nucl. Phys.}
  {\bfseries B487} (1997) 128}
  [\href{https://arxiv.org/abs/hep-th/9607210}{{\ttfamily hep-th/9607210}}].

\bibitem{Kaplan:1991dc}
D.~B. Kaplan, \emph{{Flavor at SSC energies: A New mechanism for dynamically
  generated fermion masses}},
  \href{https://doi.org/10.1016/S0550-3213(05)80021-5}{\emph{Nucl. Phys.}
  {\bfseries B365} (1991) 259}.

\bibitem{Barbieri:2007bh}
R.~Barbieri, B.~Bellazzini, V.~S. Rychkov and A.~Varagnolo, \emph{{The Higgs
  boson from an extended symmetry}},
  \href{https://doi.org/10.1103/PhysRevD.76.115008}{\emph{Phys. Rev.}
  {\bfseries D76} (2007) 115008}
  [\href{https://arxiv.org/abs/0706.0432}{{\ttfamily 0706.0432}}].

\bibitem{Vecchi:2013bja}
L.~Vecchi, \emph{{The Natural Composite Higgs}},
  \href{https://arxiv.org/abs/1304.4579}{{\ttfamily 1304.4579}}.

\bibitem{Bellazzini:2014yua}
B.~Bellazzini, C.~Csáki and J.~Serra, \emph{{Composite Higgses}},
  \href{https://doi.org/10.1140/epjc/s10052-014-2766-x}{\emph{Eur. Phys. J.}
  {\bfseries C74} (2014) 2766}
  [\href{https://arxiv.org/abs/1401.2457}{{\ttfamily 1401.2457}}].

\bibitem{Agashe:2006at}
K.~Agashe, R.~Contino, L.~Da~Rold and A.~Pomarol, \emph{{A Custodial symmetry
  for $Zb \bar b$}},
  \href{https://doi.org/10.1016/j.physletb.2006.08.005}{\emph{Phys. Lett.}
  {\bfseries B641} (2006) 62}
  [\href{https://arxiv.org/abs/hep-ph/0605341}{{\ttfamily hep-ph/0605341}}].

\bibitem{Cacciapaglia:2015dsa}
G.~Cacciapaglia, H.~Cai, T.~Flacke, S.~J. Lee, A.~Parolini and H.~Serôdio,
  \emph{{Anarchic Yukawas and top partial compositeness: the flavour of a
  successful marriage}},
  \href{https://doi.org/10.1007/JHEP06(2015)085}{\emph{JHEP} {\bfseries 06}
  (2015) 085} [\href{https://arxiv.org/abs/1501.03818}{{\ttfamily
  1501.03818}}].

\bibitem{Gripaios:2009pe}
B.~Gripaios, A.~Pomarol, F.~Riva and J.~Serra, \emph{{Beyond the Minimal
  Composite Higgs Model}},
  \href{https://doi.org/10.1088/1126-6708/2009/04/070}{\emph{JHEP} {\bfseries
  04} (2009) 070} [\href{https://arxiv.org/abs/0902.1483}{{\ttfamily
  0902.1483}}].

\bibitem{Arbey:2015exa}
A.~Arbey, G.~Cacciapaglia, H.~Cai, A.~Deandrea, S.~Le~Corre and F.~Sannino,
  \emph{{Fundamental Composite Electroweak Dynamics: Status at the LHC}},
  \href{https://doi.org/10.1103/PhysRevD.95.015028}{\emph{Phys. Rev.}
  {\bfseries D95} (2017) 015028}
  [\href{https://arxiv.org/abs/1502.04718}{{\ttfamily 1502.04718}}].

\bibitem{Backovic:2014uma}
M.~Backović, T.~Flacke, S.~J. Lee and G.~Perez, \emph{{LHC Top Partner
  Searches Beyond the 2 TeV Mass Region}},
  \href{https://doi.org/10.1007/JHEP09(2015)022}{\emph{JHEP} {\bfseries 09}
  (2015) 022} [\href{https://arxiv.org/abs/1409.0409}{{\ttfamily 1409.0409}}].

\bibitem{Bizot:2018tds}
N.~Bizot, G.~Cacciapaglia and T.~Flacke, \emph{{Common exotic decays of top
  partners}}, \href{https://doi.org/10.1007/JHEP06(2018)065}{\emph{JHEP}
  {\bfseries 06} (2018) 065}
  [\href{https://arxiv.org/abs/1803.00021}{{\ttfamily 1803.00021}}].

\bibitem{Aad:2013wta}
{\scshape ATLAS} collaboration, \emph{{Search for new phenomena in final states
  with large jet multiplicities and missing transverse momentum at $\sqrt{s}$=8
  TeV proton-proton collisions using the ATLAS experiment}},
  \href{https://doi.org/10.1007/JHEP10(2013)130,
  10.1007/JHEP01(2014)109}{\emph{JHEP} {\bfseries 10} (2013) 130}
  [\href{https://arxiv.org/abs/1308.1841}{{\ttfamily 1308.1841}}].

\bibitem{Evans:2013jna}
J.~A. Evans, Y.~Kats, D.~Shih and M.~J. Strassler, \emph{{Toward Full LHC
  Coverage of Natural Supersymmetry}},
  \href{https://doi.org/10.1007/JHEP07(2014)101}{\emph{JHEP} {\bfseries 07}
  (2014) 101} [\href{https://arxiv.org/abs/1310.5758}{{\ttfamily 1310.5758}}].

\bibitem{Buckley:2016kvr}
M.~R. Buckley, D.~Feld, S.~Macaluso, A.~Monteux and D.~Shih, \emph{{Cornering
  Natural SUSY at LHC Run II and Beyond}},
  \href{https://doi.org/10.1007/JHEP08(2017)115}{\emph{JHEP} {\bfseries 08}
  (2017) 115} [\href{https://arxiv.org/abs/1610.08059}{{\ttfamily
  1610.08059}}].

\bibitem{Tanabashi:2018oca}
{\scshape Particle Data Group} collaboration, \emph{{Review of Particle
  Physics}}, \href{https://doi.org/10.1103/PhysRevD.98.030001}{\emph{Phys.
  Rev.} {\bfseries D98} (2018) 030001}.

\bibitem{Ayyar:2018glg}
V.~Ayyar, T.~DeGrand, D.~C. Hackett, W.~I. Jay, E.~T. Neil, Y.~Shamir et~al.,
  \emph{{Partial compositeness from baryon matrix elements on the lattice}},
  \href{https://arxiv.org/abs/1812.02727}{{\ttfamily 1812.02727}}.

\end{thebibliography}\endgroup

\end{document}